\documentclass[prl,twocolumn,aps]{revtex4}
\usepackage{graphicx}

\usepackage{amsfonts}
\usepackage{amssymb}
\usepackage{amsmath}
\usepackage{hyperref}

\newcommand{\be}{\begin{equation}}
\newcommand{\ee}{\end{equation}}
\newcommand{\beq}{\begin{equation}}
\newcommand{\eeq}{\end{equation}}
\newcommand{\bea}{\begin{eqnarray}}
\newcommand{\eea}{\end{eqnarray}}

\DeclareMathOperator{\hz}{h\relax{\kern-.15em}z}
\DeclareMathOperator{\pz}{\psi\relax{\kern-.15em}z}

\begin{document}

\title{ Charged Boson Stars in AdS and a Zero Temperature Phase Transition}

\author{Sanle Hu}\email{husanle@umich.edu}
\affiliation{Department of Modern Physics, University of Science and Technology of China, Hefei, Anhui, China 230026}

\author{James T. Liu}\email{jimliu@umich.edu}
\affiliation{Michigan Center for Theoretical Physics, University of Michigan, Ann Arbor, MI 48109, USA}

\author{Leopoldo A. Pando Zayas}\email{lpandoz@umich.edu}
\affiliation{Michigan Center for Theoretical Physics, University of Michigan, Ann Arbor, MI 48109, USA}

\date{\today}
\begin{abstract}
We numerically construct charged boson stars in asymptotically AdS spacetime. We find an
intricate phase diagram dominated by solutions whose main matter contribution are
alternately provided by the scalar field or by the gauge field.
\end{abstract}

\maketitle

\noindent{\it  Introduction}\\
The study of boson stars dates back to the late 1960's with the work of Kaup \cite{Kaup:1968zz} (see also \cite{Ruffini:1969qy}) who, inspired by
ideas of Wheeler, constructed Klein-Gordon geons. These objects and their generalizations have found numerous applications in and beyond general
relativity. Some of the classical reviews in the subject include:  \cite{Lee:1991ax,Jetzer:1991jr,Liddle:1993ha,Schunck:2003kk,Liebling:2012fv}.

In this letter we construct charged boson stars in asymptotically Anti-de-Sitter (AdS) spacetimes
and investigate their properties. Namely, we
consider particle-like solutions of a complex scalar field coupled to gravity and a Maxwell field in the presence of a negative cosmological
constant. We have three main motivations to study these objects.

First, it is intrinsically interesting to understand particle-like solutions in asymptotically AdS space times to enhance and test our
intuition of highly symmetric solutions of Einstein gravity with a negative cosmological constant. For example, the role of boundary
conditions necessary to define dynamics in AdS is one aspect that constantly challenges our Minkowski-based intuition.

Our second motivation
is also linked to understanding dynamics in AdS but it is more concretely related to gravitational collapse in AdS.  It has recently been
established that AdS is unstable under arbitrarily small perturbations with respect to black hole formation \cite{Bizon:2011gg}. This result
has been discussed and elaborated upon for massless fields \cite{Jalmuzna:2011qw,Dias:2011ss,Garfinkle:2011hm,Garfinkle:2011tc,deOliveira:2012ac}.
Given previous history with critical collapse of massive \cite{Brady:1997fj} and Yang-Mills fields \cite{Choptuik:1996yg}, it is possible that the
phase diagram of critical collapse in asymptotically AdS spacetimes gets modified by the introduction of mass and other fields. In particular, the
existence of boson stars might prevent a direct channel to black hole formation.

Thirdly, solutions of a scalar field in Einstein-Maxwell gravity
in asymptotically AdS spacetimes have proven a fruitful ground for applications of the AdS/CFT correspondence to various situations in
condensed matter physics (see, for example, \cite{Hartnoll:2009sz,Horowitz:2010gk,McGreevy:2009xe}).

\noindent{\it Charged  boson stars}\\
We consider a massive charged complex scalar field, $\phi$, interacting with electromagnetism and minimally coupled to Einstein gravity with a negative cosmological constant,
\begin{eqnarray}
S&=&\int d^4x\sqrt{-g}\biggl[\frac{1}{16\pi G}(R-\Lambda)-\frac{1}{4}F_{\mu\nu}F^{\mu\nu}\nonumber \\
&&\kern6em-(D_\mu \phi)(D^\mu \phi^*)-m^2\phi^*\phi\biggr],
\label{eq:lag}\\
D_{\mu}&=&\nabla_\mu-iq\,A_\mu, \qquad F_{\mu\nu}=\nabla_{\mu}A_{\nu}-\nabla_{\nu}A_{\mu},
\end{eqnarray}
where $G$ is Newton's constant.  Since we are interested in stationary, spherically symmetric
solutions, we write the metric in Schwarzschild form
\begin{equation}
ds^2=-e^{2u}dt^2+e^{2v}d\rho^2+\rho^2d\Omega_2^2.
\end{equation}
Here $\rho$ is the radial coordinate, and we take $u=u(\rho)$ and $v=v(\rho)$. We then take
the scalar to be time-harmonic
\begin{equation}
\phi=\frac{1}{\sqrt{2}}e^{i\omega t}\sigma(\rho),
\end{equation}
where $\sigma(\rho)$ is a real function of $\rho$.  Since $\phi$ is electrically charged, it will source
the electromagnetic field, and hence we turn on a scalar potential, $A_t=A_t(\rho)$.

At this point, a few quick comments are in order.  Firstly, the Maxwell equation arising from
(\ref{eq:lag}) takes the standard form
$\nabla^\mu F_{\mu\nu}=qJ_\nu$, where
\begin{equation}
J_\mu =i(\phi^*D_\mu\phi-\phi D_\mu\phi^*)
\end{equation}
is the conserved particle number current.  As a result, the total charge of the boson star is given
by $Q=qN$ where $N$ is the conserved particle number.  Secondly, the time-dependence of $\phi$
may be removed by a gauge transformation of the form $A_t\to A_t-\omega/q$ along with
$\phi\to e^{-i\omega t}\phi$, so only the combination $\omega-qA_t$ is physical.  This is in
contrast with the standard (ungauged) boson stars, where $\omega$ has an intrinsic meaning.

It is straightforward to derive the coupled equations of motion corresponding to the above
spherically symmetric ansatz.  When developing the numerical solutions, we take as input
the mass $m$ and charge $q$ of the scalar field as well as the cosmological constant
$\Lambda$, and scale by Newton's constant $G$ when appropriate in order to work with
dimensionless quantities.  We also introduce the gauge invariant combination
\begin{equation}
\hat{A}(\rho)=(\omega-qA_t(\rho))/(\omega-qA_t(0)).
\label{eq:Ahat}
\end{equation}

To define our boundary value problem, we first demand regularity near the origin, $\rho=0$. This leads to the following conditions:
\begin{eqnarray}
u(0)&=& u_0, \quad v(0)=0, \quad  \sigma(0)=\sigma_0, \quad \sigma'(0)=0 \nonumber \\
 \quad \hat{A}(0)&=&1, \quad \hat{A}'(0)=0.
\end{eqnarray}
At asymptotic infinity, $\rho\to \infty$, AdS boundary conditions imply that the matter fields
behave as
\begin{eqnarray}
\sigma(\rho)&=& \sigma_1 \rho^{-\Delta}+ \sigma_2 \rho^{\Delta-3}, \quad  \Delta =\frac{3}{2}-\sqrt{\frac{9}{4}+(mL)^2}, \nonumber \\
A_t(\rho)&=&  a_0+a_1 \rho^{-1}.
\end{eqnarray}
Our task is to look for solutions with normalizable modes at infinity, namely $\sigma_2=0$. Therefore we have a boundary value problem which we solve numerically using shooting techniques. We determine $\sigma_2=0$ with a precision of $10^{-20}$. In our minimizing algorithm we shoot by changing $\sigma_0$ of the initial data for a given $u_0$; we use $(mL)=10$ throughout.

\noindent{\it Properties of the solutions}\\
In the absence of a cosmological constant, there is a critical charge, $q_{\rm crit}$, above
which the star does not exist in flat space. In the Newtonian limit the value of $q_{\rm crit}$ is
obtained by comparing the gravitational attraction with the electrostatic repulsion of an
elementary particle interacting with the star; the result is $Gm^2=q_{\rm crit}^2/(4\pi)$. In
the context of general relativity this result is only slightly modified.  However, for boson stars
in AdS, the value of $q_{\rm crit}$ above which a star does not exist is substantially larger.
Moreover, the phase space of solutions is quite different.

Intuitively we can understand the increase of $q_{\rm crit}$ in the presence of a negative
cosmological constant as follows. Once the electrostatic force has overcome the gravitational
pull, the only force standing in the way of charged particles flying away is the pull generated by the
cosmological constant. Therefore for large enough charges the main mechanism supporting the
star is the balance between the gauge forces and the cosmological constant.

For small charge $q$, the boson stars in AdS resemble those in flat space.  The regular
zero-node solutions have a smooth scalar profile, with maximum particle density at the core, and a
gradual fall off as a function of $\rho$.  Since the scalar field dominates the energy density, we
denote these regular solutions as {\it scalar dominated}.  For sufficiently large charge $q$, on
the other hand, we find a new type of solution where the gauge field contribution dominates the
total energy of the system.  We naturally denote these as {\it gauge field dominated}.

\begin{figure}[h!]
\resizebox{3.5in}{!}{\includegraphics{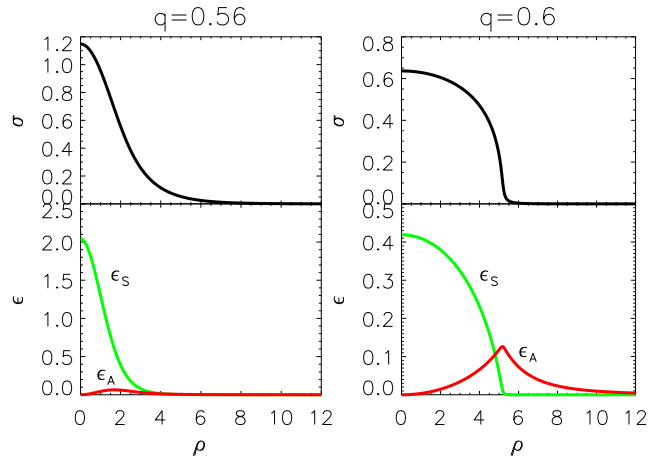}} \\
\caption{The scalar field profile $\sigma(\rho)$ (top) and energy contributions
(bottom) for regular and new solutions.  The left panel $(q=0.56)$ is for a standard charged boson
star; the right panel $(q=0.6)$ is for a new solution.}
\label{Fig.Solutions}
\end{figure}

An example of the regular and new solutions is given in Fig.~\ref{Fig.Solutions}.  The regular
scalar dominated solution is shown on the left, while the new gauge field dominated solution is on
the right. There are three key properties that distinguish the new solution from the regular one:
($i$) The new solution has a sharper surface, as defined by the profile of the scalar field; ($ii$)
The gauge field is concentrated near the surface of the star, instead of distributed
in the interior as in the standard case; ($iii$) The contribution to the total mass of the star is
dominated by the gauge field.

We now consider the mass of the boson star as a function of its core density $\sigma(0)$ and as a function of the particle number $N$.  Since the interplay between regular and new solutions depends
on the charge $q$, we define two critical charges, $q_1$ and $q_2$. For $q<q_1$ only the
scalar dominated solution is possible.  For $q_1< q< q_2$, we enter a transition region where
both the scalar dominated and gauge field dominated solutions exist as distinct branches.
Finally, for $q>q_2$, the scalar and gauge field dominated solutions merge.

\begin{figure}[h!]
\resizebox{3.5in}{!}{\includegraphics{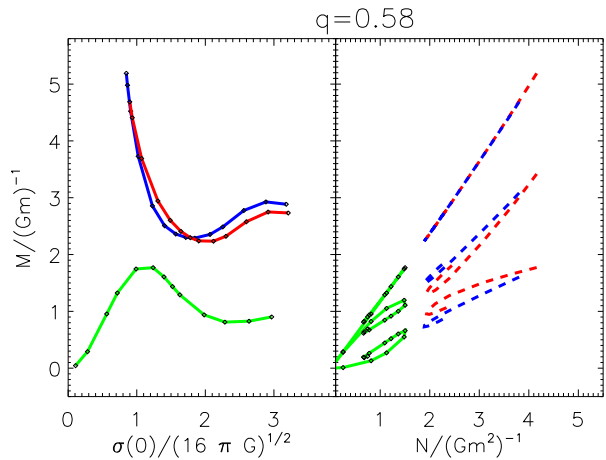}}
\caption{\label{Fig.small_q_m}Mass as a function of scalar field at origin (left) and as a function of particle number (right) in the intermediate $q$ regime.}
\end{figure}

The mass as a function of $\sigma(0)$ and as a function of $N$ for an intermediate value of
the charge, $q_1<q<q_2$, is shown in Fig.~\ref{Fig.small_q_m}. The green curve represents the
regular scalar dominated solution, while the red curve corresponds to the gauge field dominated
solution.  We have also shown the one-node solution as the blue curve.  While this is ordinarily
discarded as being an excited state, here we wish to note its proximity to the gauge field dominated
solution.  For each of these solutions, the right panel shows three curves of the same color.  These
three curves correspond to the total mass and the separate scalar and gauge field contributions
to the mass.  For the scalar dominated (green) curves, the scalar contribution to the mass is the
larger one, while for the gauge field dominated (red) curves, the gauge field contribution to the
mass is larger.  Note that there is a gap on the right panel of Fig.~\ref{Fig.small_q_m}.  This
suggests some sort of transition between scalar dominance on the left and gauge field
dominance on the right.

\begin{figure}[h!]
\resizebox{3.5in}{!}{\includegraphics{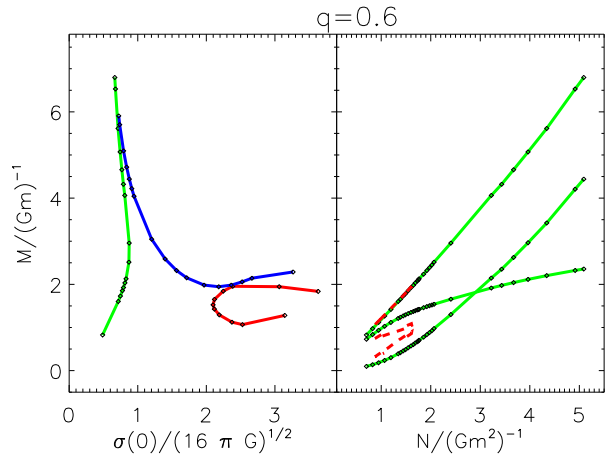}} \,\,\,\,
\caption{\label{Fig.large_q_m}Mass as a function of scalar field at origin (left) and as a function of particle number (right) in the large $q$ regime.}
\end{figure}

When the charge $q$ is increased beyond the critical value $q_2$ we observe a very different behavior of the mass as a function of $\sigma(0)$ and as a function of $N$, as shown in Fig.~\ref{Fig.large_q_m}. There is no longer a gap in the right panel and we see clearly the change in the contribution to the total energy coming from the scalar-dominated region prevalent at small values of $N$  and the gauge-dominated region at large values of $N$. The blue curve again corresponds to a one-node solution. This figure should be read as a merging of the previous one in the cases where the charge has increased, that is, increasing the charge narrows the gap between the two kind of solutions which is evident in the region of $q_1<q<q_2$.

The three green curves on the right panel of Fig.~\ref{Fig.large_q_m} are total energy,  the contribution from the scalar field and the contribution from the gauge field respectively. This graph should be read  as a merging of branches: now the green curve is a mixture of the normal branch for low $N$ and gauge-dominated branch for large $N$; the red curves are only in the areas of overlap in the previous graph. The most prominent feature of  Fig.~\ref{Fig.large_q_m} is the crossing of the scalar and gauge contributions to the total energy.

In the left panels of Figs.~\ref{Fig.small_q_m} and \ref{Fig.large_q_m} we have also plotted a branch corresponding to solutions with one node in the amplitude of the scalar field profile, $\sigma(\rho)$; these are traditionally considered to be excited solutions. What can be seen by plotting their mass is that they play an important role in the presence of a gauge field and a negative cosmological constant. In particular, the one-node branch in Fig.~\ref{Fig.large_q_m} seems to interpolate smoothly between the scalar dominated and gauge field dominated solutions. More importantly, as can be seen from Fig.~\ref{Fig.small_q_m}, the fact that the one-node and zero-node solutions come so close in terms of their energies seems to suggests that there is a mechanism taking place at large values of the charge whereby the oscillatory (usually excited) solutions come very close to the minimal energy (ground state) solution.

\begin{figure}[h!]
\begin{center}
\resizebox{3.5in}{!}{\includegraphics{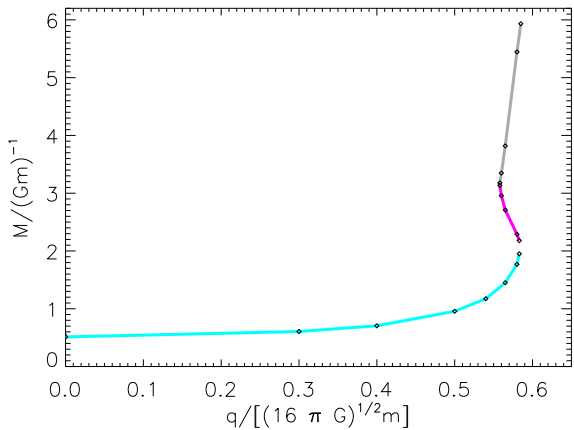}} \,\,\,\,
\caption{\label{Fig.PhaseDiagram}The phase diagram of charged boson stars in AdS.}
\end{center}
\end{figure}

\noindent {\it A Zero Temperature Phase Transition}\\
The main property of a quantum phase transition \cite{SachdevBook} is the existence of a value of a coupling $g_c$ at which there can be a level-crossing where an excited level becomes the ground state. Usually this creates a point of non-analyticity of the ground state energy as a function of $g$. Our previous graphs in Figs.~\ref{Fig.small_q_m} and \ref{Fig.large_q_m} show this non-analyticity and level crossing behavior. To emphasize this point, we plot the critical mass of a boson star as a function of charge in Fig.~\ref{Fig.PhaseDiagram}; note the previously defined $q_1$ are $q_2$ are the two inflexion points in this graph. Allowed values for boson star masses lie below and to the right of the curve.

In order to interpret Fig.~\ref{Fig.PhaseDiagram}, we note that, for a given scalar charge $q$, boson stars exist for a range values of $M$.  What we plot is the critical mass, namely the largest mass for the scalar dominated branch, and the minimum and the maximum masses for the gauge field dominated branch.   To be more graphic, the right panel in Fig.~\ref{Fig.small_q_m} contains three distinguished points: the largest mass for the green branch and the smallest and largest mass for the red branch. These are precisely the three points in the phase diagram in the region $q_1<q<q_2$.  For $q<q_1$ there is only one type of solution and we simply plot its maximum mass. Similarly for $q>q_2$, we plot only the maximum mass. In this work we are not going to be concerned with the existence of a maximum charge, $q_{\rm max}$, above which no regular solution exists. However, preliminary investigations suggest that such a point exists, and we
will discuss its determination elsewhere.

\begin{figure}[h!]
\begin{center}
\resizebox{3.5in}{!}{\includegraphics{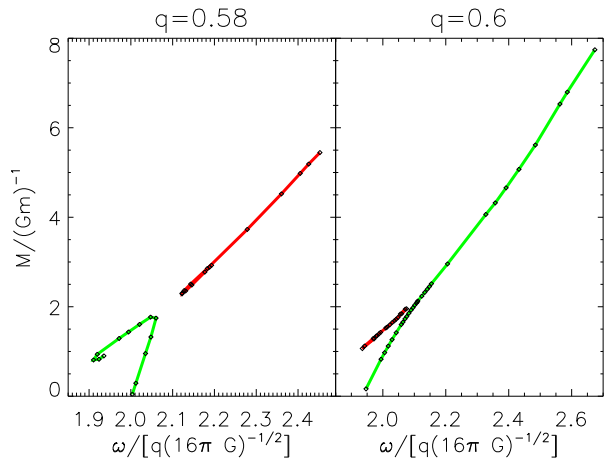}} \,\,\,\,
\caption{\label{Fig.MvOmega}The Mass as a function of the frequency $\omega$, which is a holographic proxy for chemical potential. The left panel corresponds to $q_1<q<q_2$,
while the right panel corresponds to $q>q_2$.}
\end{center}
\end{figure}

From a holographic point of view, a plot of mass versus the time component of the Maxwell
field represents the free energy as a function of chemical potential.  While we use the
gauge invariant $\hat A$ defined in (\ref{eq:Ahat}), our asymptotic boundary conditions relate
this to the value of $\omega$.  We thus plot the mass versus $\omega$ in
Fig.~\ref{Fig.MvOmega} for solutions in the intermediate and large charge regimes.

\begin{figure}[h!]
\begin{center}
\resizebox{2.5in}{!}{\includegraphics{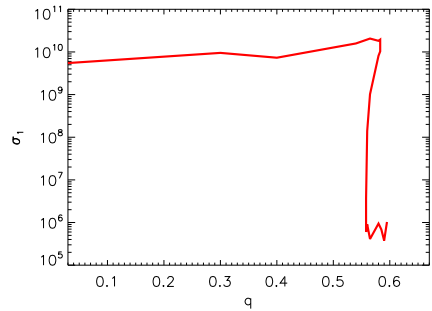}} \,\,\,\,
\caption{\label{Fig.Sigmavq} The behavior of the order parameter $\sigma_1$ as a function of the charge $q$.}
\end{center}
\end{figure}

A very important role in phase transitions is played by the order parameter. In this case we
identify it as the value of the scalar at asymptotic infinity, $\sigma_1$.  In particular, in the case
of the holographic superconductors \cite{Hartnoll:2008vx}, $\sigma_1$ corresponds to the
expectation value of an operator of mass dimension $\Delta$. In Fig.~\ref{Fig.Sigmavq} we
plot $\sigma_1$ for critical mass boson stars (those following the curve of
Fig.~\ref{Fig.PhaseDiagram}), and we see a sharp drop around the value of $q_2$.

\noindent{\it Conclusions}\\
In the context of a charged scalar field minimally coupled to Einstein-Maxwell gravity with a negative cosmological constant, we have constructed explicit solutions and established a phase diagram of boson stars. The new type of solutions contain a gauge-field-dominated branch that
represents a sort of ``Geon'' as originally envisioned by Wheeler, that is, a particle-like solution from mostly
the smooth, classical fields of electromagnetism coupled to general relativity.

Let us conclude by commenting on some open questions stemming from our work. The main difference in gravitational collapse in asymptotically AdS spaces is the presence of a timelike boundary at spatial and null infinities.  Under these conditions, the question of stability of our configurations is a particularly important one. In the phase space of initial conditions for gravitational collapse in AdS it is important to consider that a small perturbation does not escape to spatial infinity and actually returns to the center of AdS. We postpone such detailed study to a separate publication but will speculate on the role that boson stars might play.

Already in the context of asymptotically Minkowski gravitational collapse, boson stars \cite{Seidel:1991zh} play a particularly important role; they are the basis for a different type of critical collapse \cite{Brady:1997fj}. It is natural to expect that in asymptotically AdS spaces they will figure prominently.  It will be interesting to pursue the construction
and investigation of properties of real boson stars, similar to those of asymptotically Minkowski spacetimes \cite{Seidel:1991zh}, to asymptotically AdS.   One would expect that real boson stars, if they exist, get similarly destroyed after
a certain number of bounces of the scalar field at spatial infinity. Recently, however, the conjecture that boson stars might be non-linearly stable has been advanced in \cite{Dias:2012fu} based on the observation that a mass scale might prevent the resonant turbulent mechanism of \cite{Bizon:2011gg} from being realized.

Scalar boson stars have been previously considered in asymptotically AdS spacetimes by \cite{Astefanesei:2003qy,Radu:2005bp,Radu:2012yx}. Our work has concentrated on charged objects but a systematic analysis of those configurations with the corresponding interpretation and extension to the AdS/CFT applications ought to be presented. We will report on such studies, including the neutral case, in a separate publication.

By exploring the form of the boson star phase diagram, we have argued in favor of a gravity dual of a zero temperature transition for the dual field theory. It will be interesting, using the holographic dictionary, to compute explicitly various transport properties of such dual condensed matter systems.

\noindent {\it Acknowledgments}\\
We acknowledge clarifying discussions with K. Sun and C. Keeler.
S.L. is thankful to the USTC/Michigan UROP program that made possible this collaboration. L.A.P.Z. is thankful Aspen Center for Physics
for hospitality during various stages of this work.  This research was supported in part by the National Science Foundation under Grant No. NSF PHY11-25915 (KITP),  grant No. 1066293 (Aspen) and by Department of Energy under grant DE-SC0007859 to the University of Michigan.

\bibliographystyle{JHEP}
\bibliography{Collapsebib}

\providecommand{\href}[2]{#2}\begingroup\raggedright\begin{thebibliography}{10}

\bibitem{Kaup:1968zz}
D.~J. Kaup, {\it {Klein-Gordon Geon}},  {\em Phys.Rev.} {\bf 172} (1968)
  1331--1342.

\bibitem{Ruffini:1969qy}
R.~Ruffini and S.~Bonazzola, {\it {Systems of selfgravitating particles in
  general relativity and the concept of an equation of state}},  {\em
  Phys.Rev.} {\bf 187} (1969) 1767--1783.

\bibitem{Lee:1991ax}
T.~Lee and Y.~Pang, {\it {Nontopological solitons}},  {\em Phys.Rept.} {\bf
  221} (1992) 251--350.

\bibitem{Jetzer:1991jr}
P.~Jetzer, {\it {Boson stars}},  {\em Phys.Rept.} {\bf 220} (1992) 163--227.

\bibitem{Liddle:1993ha}
A.~R. Liddle and M.~S. Madsen, {\it {The Structure and formation of boson
  stars}},  {\em Int.J.Mod.Phys.} {\bf D1} (1992) 101--144.

\bibitem{Schunck:2003kk}
F.~Schunck and E.~Mielke, {\it {General relativistic boson stars}},  {\em
  Class.Quant.Grav.} {\bf 20} (2003) R301--R356,
  [\href{http://xxx.lanl.gov/abs/0801.0307}{{\tt arXiv:0801.0307}}].

\bibitem{Liebling:2012fv}
S.~L. Liebling and C.~Palenzuela, {\it {Dynamical Boson Stars}},  {\em Living
  Rev.Rel.} {\bf 15} (2012) 6, [\href{http://xxx.lanl.gov/abs/1202.5809}{{\tt
  arXiv:1202.5809}}].

\bibitem{Bizon:2011gg}
P.~Bizon and A.~Rostworowski, {\it {On weakly turbulent instability of anti-de
  Sitter space}},  {\em Phys.Rev.Lett.} {\bf 107} (2011) 031102,
  [\href{http://xxx.lanl.gov/abs/1104.3702}{{\tt arXiv:1104.3702}}].

\bibitem{Jalmuzna:2011qw}
J.~Jalmuzna, A.~Rostworowski, and P.~Bizon, {\it {A Comment on AdS collapse of
  a scalar field in higher dimensions}},  {\em Phys.Rev.} {\bf D84} (2011)
  085021, [\href{http://xxx.lanl.gov/abs/1108.4539}{{\tt arXiv:1108.4539}}]. 3
  pages, 2 figures.

\bibitem{Dias:2011ss}
O.~J. Dias, G.~T. Horowitz, and J.~E. Santos, {\it {Gravitational Turbulent
  Instability of Anti-de Sitter Space}},
  \href{http://xxx.lanl.gov/abs/1109.1825}{{\tt arXiv:1109.1825}}. * Temporary
  entry *.

\bibitem{Garfinkle:2011hm}
D.~Garfinkle and L.~A. Pando~Zayas, {\it {Rapid Thermalization in Field Theory
  from Gravitational Collapse}},  {\em Phys.Rev.} {\bf D84} (2011) 066006,
  [\href{http://xxx.lanl.gov/abs/1106.2339}{{\tt arXiv:1106.2339}}].

\bibitem{Garfinkle:2011tc}
D.~Garfinkle, L.~A. Pando~Zayas, and D.~Reichmann, {\it {On Field Theory
  Thermalization from Gravitational Collapse}},  {\em JHEP} {\bf 1202} (2012)
  119, [\href{http://xxx.lanl.gov/abs/1110.5823}{{\tt arXiv:1110.5823}}].

\bibitem{deOliveira:2012ac}
H.~de~Oliveira, L.~A. Pando~Zayas, and C.~A. Terrero-Escalante, {\it
  {Turbulence and Chaos in Anti-de-Sitter Gravity}},
  \href{http://xxx.lanl.gov/abs/1205.3232}{{\tt arXiv:1205.3232}}.

\bibitem{Brady:1997fj}
P.~R. Brady, C.~M. Chambers, and S.~M. Goncalves, {\it {Phases of massive
  scalar field collapse}},  {\em Phys.Rev.} {\bf D56} (1997) 6057--6061,
  [\href{http://xxx.lanl.gov/abs/gr-qc/9709014}{{\tt gr-qc/9709014}}].

\bibitem{Choptuik:1996yg}
M.~W. Choptuik, T.~Chmaj, and P.~Bizon, {\it {Critical behavior in
  gravitational collapse of a Yang-Mills field}},  {\em Phys.Rev.Lett.} {\bf
  77} (1996) 424--427, [\href{http://xxx.lanl.gov/abs/gr-qc/9603051}{{\tt
  gr-qc/9603051}}].

\bibitem{Hartnoll:2009sz}
S.~A. Hartnoll, {\it {Lectures on holographic methods for condensed matter
  physics}},  {\em Class.Quant.Grav.} {\bf 26} (2009) 224002,
  [\href{http://xxx.lanl.gov/abs/0903.3246}{{\tt arXiv:0903.3246}}].

\bibitem{Horowitz:2010gk}
G.~T. Horowitz, {\it {Introduction to Holographic Superconductors}},
  \href{http://xxx.lanl.gov/abs/1002.1722}{{\tt arXiv:1002.1722}}.

\bibitem{McGreevy:2009xe}
J.~McGreevy, {\it {Holographic duality with a view toward many-body physics}},
  {\em Adv.High Energy Phys.} {\bf 2010} (2010) 723105,
  [\href{http://xxx.lanl.gov/abs/0909.0518}{{\tt arXiv:0909.0518}}].

\bibitem{SachdevBook}
S.~Sachdev, {\it {Quantum Phase Transitions}}, . Cambridge University Press,
  (2011) 501p.

\bibitem{Hartnoll:2008vx}
S.~A. Hartnoll, C.~P. Herzog, and G.~T. Horowitz, {\it {Building a Holographic
  Superconductor}},  {\em Phys. Rev. Lett.} {\bf 101} (2008) 031601,
  [\href{http://xxx.lanl.gov/abs/0803.3295}{{\tt arXiv:0803.3295}}].

\bibitem{Seidel:1991zh}
E.~Seidel and W.~Suen, {\it {Oscillating soliton stars}},  {\em Phys.Rev.Lett.}
  {\bf 66} (1991) 1659--1662.

\bibitem{Dias:2012fu}
O.~J. Dias, G.~T. Horowitz, D.~Marolf, and J.~E. Santos, {\it {On the Nonlinear
  Stability of Asymptotically Anti-de Sitter Solutions}},
  \href{http://xxx.lanl.gov/abs/1208.5772}{{\tt arXiv:1208.5772}}.

\bibitem{Astefanesei:2003qy}
D.~Astefanesei and E.~Radu, {\it {Boson stars with negative cosmological
  constant}},  {\em Nucl.Phys.} {\bf B665} (2003) 594--622,
  [\href{http://xxx.lanl.gov/abs/gr-qc/0309131}{{\tt gr-qc/0309131}}].

\bibitem{Radu:2005bp}
E.~Radu and E.~Winstanley, {\it {Conformally coupled scalar solitons and black
  holes with negative cosmological constant}},  {\em Phys.Rev.} {\bf D72}
  (2005) 024017, [\href{http://xxx.lanl.gov/abs/gr-qc/0503095}{{\tt
  gr-qc/0503095}}].

\bibitem{Radu:2012yx}
E.~Radu and B.~Subagyo, {\it {Spinning scalar solitons in anti-de Sitter
  spacetime}},  \href{http://xxx.lanl.gov/abs/1207.3715}{{\tt
  arXiv:1207.3715}}.

\end{thebibliography}\endgroup

\end{document}